\documentstyle[12pt]{article}
\newcommand{\Eqref}[1]{Eq.(\ref{#1})}
\newcommand{\Eqsref}[2]{Eqs.(\ref{#1},\,\ref{#2})}

\newcommand{\bea}{\begin{eqnarray}}
\newcommand{\eea}{\end{eqnarray}}
\newcommand{\be}{\begin{equation}}
\newcommand{\ee}{\end{equation}}
\newcommand{\bc}{\begin{center}}
\newcommand{\ec}{\end{center}}
\newcommand{\ba}{\begin{array}}
\newcommand{\ea}{\end{array}}
\newcommand{\bfig}{\begin{figure}}
\newcommand{\efig}{\end{figure}}
\newcommand{\non}{\nonumber}
\newcommand{\lefq}[1]{\lefteqn{#1}}
\newcommand{\inv}[1]{\frac{1}{#1}}
\newcommand{\del}{\partial}

\newcommand{\Abs}[1]{\left|#1\right|}

\newcommand{\ra}{\rangle}
\newcommand{\la}{\langle}

\newcommand{\vek}[1]{{\rm\bf}{#1}}

\newcommand{\prd}[3]{{\it  Phys. Rev. D} {{\bf #1} {(#2)} {#3}}}

\newcommand{\pl}[3]{{\it  Phys. Lett. }{{\bf #1} {(#2)} {#3}}}

\newcommand{\zp}[3]{{\it Z. Phys.} {{\bf #1} {(#2)} {#3}}}

\newcommand{\etal}{\mbox{\sl et al. }}
\newcommand{\PE}{\ \\ {\centering {\large Per Elmfors} \\ \ \\
   { NORDITA \\
   Blegdamsvej 17 \\
   DK-2100 K\o benhavn \O \\
   Denmark \\
   E-mail: elmfors@nordita.dk \\ }}}

\newcommand{\phits}{(\phi^3)_6}
\newcommand{\phiff}{(\phi^4)_4}
\newcommand{\me}{\mu^{\epsilon}}
\newcommand{\mte}{\mu^{2\epsilon}}
\newcommand{\mme}{\mu^{-\epsilon}}
\newcommand{\mmte}{\mu^{-2\epsilon}}
\newcommand{\tr}{T_R}
\newcommand{\mr}{m^2_R}
\newcommand{\gr}{g_R}
\newcommand{\lr}{\lambda_R}
\newcommand{\sr}{\sigma_R}
\newcommand{\mmr}{M^2_R}
\newcommand{\lst}{\lambda\sigma^2}
\newcommand{\lstr}{\lambda_R\sigma^2_R}

\normalsize
\begin{document}
\thispagestyle{empty}
\begin{flushright} NORDITA-92/38  P \\
                   hep-ph/9206240  \\
                   May 1992  \end{flushright}
\bc
{\LARGE\bf Finite Temperature Renormalization\\
 of the $(\phi^3)_6$- and $(\phi^4)_4$-Models\\
 at Zero Momentum}
\ec
\vspace*{1cm}
\PE
\vspace*{6cm}
\nopagebreak
\begin{abstract}
A self-consistent renormalization scheme at finite temperature and
zero momentum is used
together with the finite temperature renormalization group to study
the temperature dependence of the mass and the coupling to one-loop
order in the
$(\phi^3)_6$- and $(\phi^4)_4$-models. It is found that the
critical temperature is shifted relative to the naive one-loop
result and the coupling constants at the critical temperature get
large corrections. In the high temperature limit of the $\phiff$-model
the coupling decreases.
\end{abstract}
\newpage
%
\setcounter{page}{1}
\section{Introduction}
Relativistic finite temperature field theory is an important tool
when studying phase transitions in the early universe. A detailed
understanding of the electroweak phase transition is needed to
determine limits of the Higgs mass and possibly rule out the
simplest version of the standard model as a theory of
electroweak baryogenesis \cite{dinelhll92}. In this case, as in many
others, infrared (IR) problems plague the perturbation theory since
factors of $T/M$ ($T$ is the temperature and $M$ a typical mass)
enter with higher powers in higher loops. Large $T$ singularities
are thus similar to small $M$ singularities in this respect.
One way to make the divergence weaker is to sum over an infinite
set of diagrams and thereby give a thermal contribution to the mass.
In the $(\phi^4)_4$-theory, where the one-loop correction to the mass
is momentum independent, such a resummation gives only a $T$ dependent
shift of the mass. At large $T$ the mass, corrected in this way, is
\be
\label{m2=t2}
      M^2(T)=m^2+\Sigma_\beta(m^2,\lambda,T)\simeq
      m^2+\frac{\lambda T^2}{24}\ ,
\ee
where $\Sigma_\beta$ is the finite temperature part of the
self-energy and $m$ is the mass renormalized
at zero temperature. When this summation is used iteratively
on all internal lines a  gap equation is obtained
containing all {\it superdaisy}
diagrams in the language of Ref.\cite{dolanj74}
\be
\label{gap}
      M^2(T)=m^2+\Sigma_\beta(M^2(T),\lambda,T)\ .
\ee
This type of resummation, or a first iteration of it, was
recently used for the standard electroweak theory
 \cite{carrington91}.
 A condition for being able to
perform the resummation is that the correction is momentum
independent, otherwise one gets complicated integral equations
(Schwinger-Dyson equations). Such a simple resummation
 is not possible to carry out
for the coupling constant though it may also get large
corrections in the IR limit. Its one-loop
correction has the following high $T$ behaviour
\be
      \lambda(T)\simeq\lambda-\frac{3\lambda^2}{16\pi}
      \frac{T}{M}\ .
\ee
\\ \\
In the IR limit the corrections to mass and coupling are
large and the perturbation theory in terms of the zero temperature
parameters breaks down. This problem can be circumvented if the
renormalization is performed at the temperature in which one
interested, so that the mass and coupling take the physical values.
Such
 a renormalization condition
absorbs a dynamical change of the parameters into a constant at a given
renormalization point. In next section this idea is further supplemented by
the use of the temperature renormalization group equation,
as derived by Matsumoto \etal \cite{matsumotonu84}, to get the
$T$ dependence of the renormalized parameters.
\\ \\
The IR singularities occur not only in the high
temperature limit but also at a second order phase transition
where the mass becomes zero at finite $T$. The phase
transitions in $(\phi^3)_6$- and $(\phi^4)_4$-models are studied
in section \ref{phitre} and \ref{phifyra}.
%
\section{Renormalization at finite temperature}
\label{renatfint}
It is often emphasized that finite temperature does not introduce new
UV divergences and that it is, therefore, enough to renormalize at
zero temperature. The physical quantities should anyway be independent
of the renormalization scheme. However, to finite order in
perturbation theory the physical quantities do indeed depend
on finite changes of the renormalization prescription and it
is necessary to choose the prescription carefully. QCD is an example
where this is important and the running of the coupling constant with
temperature depends crucially on the vertex chosen for the
renormalization \cite{fujimotoy88b,nakkagawany88}.
One strategy for renormalization
is that as much as possible of the dynamics should be put into
the expansion parameters. The same philosophy was discussed in
Ref.\cite{freires91}, though their treatment of the problem was
slightly different from the one in this paper.

In the usual renormalization group approach
the mass and the coupling in the
Feynman rules are chosen to coincide with the measured values at
some relevant scale $\mu$.
Perturbation theory is then used to compute the mass and the
coupling at another scale $\mu'$.  If the difference between
$\mu$ and $\mu'$ is small one can hope that perturbation theory
is good. Finally, the renormalization group is obtained as
a differential equation in the limit $\mu'\rightarrow\mu$.
The aim of this paper is to use this idea at
finite temperature to the $\phits$- and
$\phiff$-models below the critical temperature.

The  temperature renormalization
group equations ((T)RGE) were introduced by Matsumoto \etal
\cite{matsumotonu84}. It has been pointed out that a naive
analogue of the usual RGE is not valid since physical quantities
do depend on the temperature. But that is not the point of the
TRGE derived in \cite{matsumotonu84}. There, it is only used that the
physical quantities should be independent of the temperature
at which one chooses to renormalize. This will then
determine the $T$ dependence of the physical quantities.
\\ \\
An intuitive understanding of the TRGE can be gained in the
following way. Suppose we renormalize the Lagrangian at a
temperature $T$ and then compute the mass at $T+\Delta T$.
We then get
\be
      M^2(T+\Delta T)=M^2(T)+\Sigma(M^2(T),T+\Delta T)
      - \Sigma(M^2(T),T) \ ,
\ee
where $\Sigma$ is the self-energy. If the difference between
$\Sigma(T+\Delta T)$ and $\Sigma(T)$ is large we cannot trust
perturbation theory but in the limit $\Delta T\rightarrow 0$ it is
expected to be reliable. Therefore we instead derive
\be
\label{dmdt}
      \frac{d M^2}{d T}=\frac{\del \Sigma(M^2,T)}
      {\del T} \ ,
\ee
and solve this differential equation.\footnote{
Note that the gap equation (\Eqref{gap}) gives
$\frac{d M^2}{d T}=\frac{d\Sigma(M^2(T),T)}
      {dT}\ ,$ with total derivatives.}
In each step of the integration
of \Eqref{dmdt} the correction is small and we are more likely to
stay within
the range of validity of perturbation theory.
\\ \\
A derivation of the TRGE for an unbroken theory
can be found in Ref.\cite{matsumotonu84}.
When the theory is spontaneously broken the renormalization condition
(RC) has to be changed in order for the mass in the propagator
to be the physical mass. Let us start from  a bare Lagrangian
$L=L(\phi_0,m^2_0,g_0)$ where the mass $m^2_0$ and the $k$-point
coupling $g_0$ are fixed constants. They are temperature
independent but they have to be infinite to make physical quantities
finite in
renormalizable theories . We
choose to use dimensional regularization so that $m^2_0$ and $g_0$
depend on $1/\epsilon$. Next we rescale the field $\phi_0=\phi Z^{1/2}$
so the expectation value of $\phi$ is finite. This can be considered
as a renormalization of $\hbar$.

Then we divide $L$ into a finite part with which we define the
perturbation theory and the rest is considered as counterterms.
Typically we have
\be
      \frac{m^2_0}{2}\phi^2_0=\frac{m^2_0Z}{2}\phi^2=
      \frac{m^2}{2}\phi^2+\inv{2}(m^2_0Z-m^2)\phi^2 \ ,
\ee
and similarly for $g_0\rightarrow g$. This does not change $L$
and everything computed from $L$ is formally independent of $m^2$ and $g$.
Now we determine $m^2_0$, $g_0$ and $Z$ from the measured values
of the mass and coupling at a given temperature $\tr$. The physical
mass and coupling ($\mmr$ and $\gr$) are related to the excitations
around the actual minimum (which may be different from $\phi=0$)
so we shift the field $\phi\rightarrow \phi+\sigma$ and use the
following RC (called the $\tr$-scheme with the terminology
of Ref.\cite{matsumotonu84})
\bea
\label{RC1}
      {\rm Re}\Gamma^{(1)}(\tr,\sr,\mmr,\gr) &=& 0 \\
\label{RC2}
      {\rm Re}\Gamma^{(2)}(p(Q_R),\tr,\sr,\mmr,\gr) &=&  p^2 (Q_R)-\mmr(T_R) \\
\label{RC3}
      \frac{\del}{\del p^2}{\rm Re}\Gamma^{(2)}(p(Q_R),\tr,\sr,\mmr,\gr)
      &=&   1 \\
\label{RC4}
      {\rm Re}\Gamma^{(k)}(p_i(Q_R),\tr,\sr,\mmr,\gr) &=& -\gr(T_R)\ ,
\eea
where $\Gamma^{(N)}$ are the $N$-point functions in the effective
action computed from the shifted Lagrangian using $\mmr$ as mass and
$\gr$ as coupling in the perturbation theory. The momentum is chosen
in some suitable way (see the discussion in {\mbox section \ref{phitre})}.
At finite temperature relaxation processes introduce imaginary parts
in $\Gamma^{(N)}$ so we take the real part in the RC.
The first condition ensures that the effective action is expanded
around a stationary point (a minimum should be chosen). These conditions
determine $m^2_0$, $g_0$, $Z$ and $\sr$ in terms of $\mmr$, $\gr$, $\tr$
and $Q_R$ (and the scale of dimensional regularization $\mu$ which
is not spelled out). We can then write down the Lagrangian with
unshifted fields in terms of the renormalized quantities and
counterterms which depend on $1/\epsilon$. From this Lagrangian
we compute finite $N$-point functions and denote them by $\Gamma^{(N)}
(p,T,\sigma;\sr,\mmr,\gr,\tr)$.
\\ \\
A change in renormalization temperature to $t\tr$ means that
$\sr$, $\mmr$ and $\gr$ are different, but the final $N$-point
function must be the same up to a rescaling of the field:
\bea
\label{tscale}
\lefteqn{
      \Gamma^{(N)}(p,T,\sigma;\tr,\sr(\tr),\mmr(\tr),
      \gr(\tr))}\non\\
      &= &\rho(t\tr)^{-N/2}
      \Gamma^{(N)}
      (p,T,\sigma;t\tr,\sr(t\tr),\mmr(t\tr),\gr(t\tr))\ ,
\eea
where $\rho$ is the finite wavefunction renormalization.
This invariance imply
\be
\label{TRGE}
      \left( \tr\frac{\del}{\del\tr}+\eta\frac{\del}{\del\sr}
      +\theta M_R\frac{\del}{\del M_R}+\beta\frac{\del}{\del\gr}
      -N\gamma\right)\Gamma^{(N)}
      (p,T,\sigma;\tr,\sr,\mmr,\gr)=0\ ,
\ee
where
\be
\label{RGFdef}
      \eta=\tr\frac{d\sr}{d\tr}\ , \
      \theta=\frac{\tr}{M_R}\frac{d M_R}{d\tr} \ ,\
      \beta=\tr\frac{d\gr}{d\tr} \ ,\
      \gamma=\frac{\tr}{2\rho}\frac{d\rho}{d\tr} \ .
\ee
Let us abbreviate the notation and write
\be
      \Gamma^{(N)}(p,T,\sigma)=
      \Gamma^{(N)}(p,T,\sigma;\tr,\sr(\tr),\mmr(\tr),\gr(\tr))\ .
\ee
Using \Eqref{RC1}-\Eqref{RC4} and \Eqref{tscale} one can derive
the TRGE
(see Ref.\cite{matsumotonu84})
\bea
\label{renfct}
      0&=&(T\frac{\del}{\del T} +\eta(\tr)\frac{\del}{\del\sigma})
      {\rm Re}\Gamma^{(1)}(T,\sigma)|_{\tr,\sr}\ , \non\\
      \theta(\tr)&=&\gamma\frac{\mmr-p^2(Q_R)}{\mmr}
      -\inv{2\mmr}
      \left(T\frac{\del}{\del T}+\eta\frac{\del}{\del\sigma}\right)
      {\rm Re}\Gamma^{(2)}(p,T,\sigma)|_{p(Q_R),\tr,\sr}\ , \non\\
      \gamma(\tr)&=&-\inv{2}
      \left(T\frac{\del}{\del T}+\eta\frac{\del}{\del\sigma}\right)
      \frac{\del}{\del p^2}
      {\rm Re}\Gamma^{(2)}(p,T,\sigma)|_{p(Q_R),\tr,\sr}\ ,\non\\
      \beta(\tr)&=&kg\gamma -
      \left(T\frac{\del}{\del T}+\eta\frac{\del}{\del\sigma}\right)
      {\rm Re}\Gamma^{(k)}(p_i,T,\sigma)|_{p_i(Q_R),\tr,\sr}\ .
\eea
Notice that the $T$ and $\sigma$ derivatives only act upon the explicit $T$
and $\sigma$
dependence. The right hand side can be computed perturbatively in
the coupling and then one is left with a set of coupled differential
equations which defines a mass and a coupling {\it running} with
the temperature.
\\ \\
The equations above are written in a Lorentz invariant way but
the actual vertices do not respect the Lorentz symmetry because of the
thermal heat-bath. The wavefunction renormalization condition
has to be modified in some way and I use $\frac{\del}{\del p^2_0}$
instead of $\frac{\del}{\del p^2}$. The momentum point can be chosen
to be $p(Q_R)= (Q_R,0,0,0)$.
\\ \\
The effective potential can be calculated from the 1-point function
of the shifted theory using
\be
      \frac{dV(T,\sigma)}{d\sigma}=-\Gamma^{(1)}(T,\sigma)\ .
\ee
When $\Gamma^{(1)}$ is expressed in terms of the renormalized quantities
$\sr$, $\mmr$ and $\gr$ all the infinities ($\propto 1/\epsilon$)
and the $\mu$ dependence disappear.
%
%
\section{The $(\phi^3)_6$-model}
\label{phitre}
The scalar $(\phi^3)_6$-model, being renormalizable and asymptotically free,
has been investigated as a toy model
of QCD. The
potential is unbounded from below but for small coupling and
positive mass it has a local minimum. At finite temperature it
is expected that the decay-rate increases and that the system becomes
completely unstable above a critical temperature. This was verified
in Ref.\cite{altherrgp91} where the critical temperature was
found to be
\be
      T_{cr}=\left(\frac{180}{\pi}\right)^{1/4}\frac{M}{\sqrt{g}}\ .
\ee
Here $M$ and $g$ are the zero temperature mass and coupling constant.

The expression for the effective mass derived in \cite{altherrgp91}
can be obtained using the naive TRGE in section \ref{renatfint}. The
minimum of the effective potential
is at $\la \phi\ra=-\frac{g\pi T^4}{360M^2}$ in the high
temperature limit (but below the critical point) and there the
effective $T$ dependent mass is
\be
      M^2(T)=M^2+g\la\phi\ra=M^2-\frac{g^2\pi T^4}{360M^2}\ .
\ee
This is a one-loop correction which is a polynomial in $g$. If we
instead solve the differential equation in \Eqref{dmdt} we get
\be
      M^2(T)=M^2\sqrt{1-\frac{g^2\pi T^4}{180M^4}}\ ,
\ee
which corresponds to the sum of an infinite set of diagrams with
leading power in temperature (see Ref.\cite{altherrgp91}) and it is
no longer a polynomial in $g$.
\\ \\
Let us now use the $\phits$-theory as a simple model to see
how the renormalization procedure described in section \ref{renatfint}
works out in practise.

The Lagrangian in $D=6-2\epsilon$ dimensions is
\bea
\label{lag3}
\lefteqn{
      L=\inv{2}(\del_\mu\phi_0)^2-h_0\phi_0-
      \frac{m_0^2}{2}\phi_0^2 - \frac{g_0}{6}\phi^3_0 }\non\\
     &=& \inv{2}(\del_\mu\phi)^2-\frac{M^2}{2}\phi^2
      - \frac{g\me}{6}\phi^3
       +\frac{Z-1}{2}(\del_\mu\phi)^2
      - h_0Z^{1/2}\phi-(m^2_0Z+\frac{g_0Z^{3/2}\sigma\mme}{2})
      \sigma\mme\phi \non\\
      & &- \inv{2}(m^2_0Z+g_0Z^{3/2}\sigma\mme-M^2)
      \phi^2 -\frac{g_0Z^{3/2}-g\me}{6}\phi^3 \ .
\eea
where $Z$, $\sigma$, $M^2=m^2+g\sigma$ and $g$ are
arbitrary parameters and a
change of them is a change of the renormalization prescription. To
finite order (in $g$) the result depends on how we have chosen them.
The Lagrangian above is obtained by scaling the field ($\phi_0=Z\phi'$),
separating the parameters into finite and infinite parts
($m^2_0Z=m^2+(m^2_0Z-m^2)$ etc.) and finally shifting the field
($\phi'= \phi+\sigma\mme$). The mass scale $\mu$ is introduced
to give $g$ and $\sigma$ the same dimensions as in $D=6$.
\\ \\
For simplicity I compute
wavefunction renormalization for non-zero external momentum
 but let then the momentum go to zero.
All other $N$-point functions are computed by taking derivatives
of the 1-point function with respect to $\sigma$, i.e. at zero momentum.
It is, after all, the IR limit ($p\rightarrow 0$)
that is of most concern and that
is presumably taken care of with this  prescription.
One could, on the other hand, also argue that $p$ should be of
order $T$ since the particles in the gas typically have that
momentum \cite{enqvistk92}. At the critical point, where the
temperature is finite but the mass goes to zero, a finite $p$ could
act as a regulator.

In connection with this I would also like the remind the reader
that the limit of taking the external momentum to zero is
non-trivial at finite temperature. As was shown by Fujimoto \etal
\cite{fujimotoy88} the limit depends on whether the momentum
is space-like ($p^2<0$) or time-like ($p^2>0$). The calculation
in Ref.\cite{fujimotoy88}
was performed in the real-time formalism and coincidence with
the imaginary time formalism was only found in the space-like
limit. It does however not tell which limit is the correct one. I
stick to the space-like limit for  simplicity except for the
wavefunction renormalization.

The easiest way to compute the effective potential, from which
we can derive the $N$-point vertices at zero external momenta, is to
calculate the tadpole diagram for the shifted Lagrangian in \Eqref{lag3}.
The tadpole is equal to the derivative of the
effective potential with respect
to the field up to a factor $i$.
The result is the following
\bea
\lefq{
      \Gamma^{(1)}(T,\sigma,M,g)
      =-h_0Z^{1/2}-(m^2_0Z+\frac{g_0Z^{3/2}\sigma\mme}{2})\sigma\mme}
      \non\\
      & &-\frac{g\mme M^4}{256\pi^3}\left(\inv{\epsilon}+\frac{3}{2}
      -\gamma-\ln(\frac{M^2}{4\pi\mu^2})\right)
      -\frac{g}{24\pi^3}F^4_1(T,M^2)\ ,  \\
\lefq{
      \Gamma^{(2)}(p=0,T,\sigma,M,g)=
      -(m^2_0Z+g_0Z^{3/2}\sigma\mme)}\non\\
      & &-\frac{g^2M^2}{128\pi^3}\left(\inv{\epsilon}
      +1-\gamma-\ln(\frac{M^2}
      {4\pi\mu^2})\right) + \frac{g^2}{16\pi^3}F^2_1(T,M^2) \ ,\\
\lefq{
      \frac{\del}{\del p^2_0}\Gamma^{(2)}(p=0,T,\sigma,m,g)}\non\\
      &=&Z+\frac{g^2}{768\pi^3}\left(\inv{\epsilon}-\gamma
      -\ln(\frac{M^2}{4\pi\mu^2})\right)
      +\frac{g^2}{192\pi^3}F^4_5(T,M^2)\ , \\
\lefq{
      \Gamma^{(3)}(p_i=0,T,\sigma,m,g)}\non\\
      &=&-g_0Z^{3/2}-\frac{g^3\me}{128\pi^3}\left(\inv{\epsilon}-\gamma
      -\ln(\frac{M^2}{4\pi\mu^2})\right)
      -\frac{g^3}{32\pi^3}F^0_1(T,M^2) \ .
\eea
We have defined some useful functions
\be
      F^m_n(T,M^2)=\int_0^\infty dk\frac{k^mf_B(\omega)}{\omega^n}
      \ , \ f_B(\omega)=\inv{e^{\omega/T}-1} \ , \
      \omega=\sqrt{\vek{k}^2+M^2}\ .
\ee
\\ \\
The RC in \Eqref{RC1}-\Eqref{RC4} are then used at $\tr$
to determine the infinite constants
$h_0$, $m^2_0$, $g_0$ and $Z$ ($h_0$ and $\sigma$ are not independent
but we fix $h_0$ at zero temperature where $\sigma=0$). After that
we compute $\Gamma^{(N)}(T,\sigma)$ and express it terms of finite
quantities. When we first fix $\mmr$ at a shifted field $\sr$, shift
the field
back again to get the original Lagrangian and finally shift with an
arbitrary $\sigma$ to compute the effective potential from the tadpole,
it effectively equals fixing $\mr=\mmr-\gr\sr$ instead of $\mmr$. So
in the expression for $\Gamma^{(1)}(T,\sigma)$ there is a $\sigma$
dependence in $M^2=\mr+\gr\sigma$.

The effective potential can be integrated explicitly from
$\Gamma^{(1)}(T,\sigma)$ and we find
\bea
\label{V3}
\lefq{
      V(T,\sigma)= -(\sigma-\sr)\left(\frac{\gr M^4_R}{512\pi^3}+
      \frac{\gr}{24\pi^3}F^4_1(\tr,\mmr)\right) }\non\\
      & &+\frac{(\sigma-\sr)^2}{2}\left(\mmr+\frac{\gr^2}{16\pi^3}
      F^2_1(\tr,\mmr)
      \right) \non\\
      & &+\frac{(\sigma-\sr)^3}{6}\left(\gr+\frac{\gr^3}{128\pi^3}
      -\frac{\gr^3}{32\pi^3}F^0_1(\tr,\mmr)
      \right) \non\\
      & &+\frac{M^6}{768\pi^3}\left(\frac{5}{6}-\ln(\frac{M^2}{\mmr})
      \right) - \inv{60\pi^3}F^6_1(T,M^2)+{\rm const.}\ ,
\eea
where we have used
\be
      \frac{\del F^m_n}{\del M^2}=-\frac{m-1}{2}F^{m-2}_n
      \ ,\ m \ge 2 \ .
\ee
If we choose the renormalization temperature to be equal to
the actual temperature we get an effective action which
has the values of the mass and the coupling at the minimum
coinciding with the values of the {\it running} mass
and coupling from the TRGE. Also, at $T=\tr$ the minimum is
at $\sigma=\sr$.

The form of the effective potential as a function of $\sigma$
is essentially the same as the zero temperature renormalization
scheme but now the parameters have a non-trivial dependence of
the temperature $\tr$. Therefore, we still have a minimum which
becomes shallower  as the temperature increases and
eventually disappears. The value of the critical temperature
is however different and depends on the zero temperature
mass and coupling  in a non-trivial way.
\\ \\
The $\tr$ dependence of $\mr$ and $\gr$ should also be calculated.
By taking derivatives of $V(T,\sigma)$ with respect to $\sigma$
we get finite expressions for the other $\Gamma^{(N)}$'s at
zero momentum, and they are used to get the one-loop TRGE for
the $(\phi^3)_6$-theory. Dropping the subscript $R$ we get:
\bea
      0&=&\frac{g}{24\pi^3}\frac{\del F^4_1}{\del T} +
      \frac{d\sigma}{dT} M^2 \\
      \frac{\gamma}{T}&=&-\frac{g^2}{384\pi^3}
      \left(\frac{\del F^4_5}{\del T}
      -g\frac{d\sigma}{d T}(\inv{4M^2}-
      \frac{\del F^4_5}{\del M^2})\right)\\
      \frac{dM}{dT}&=&M\frac{\gamma}{T}+\frac{g}{2M}\frac{d\sigma}{dT}
      -\frac{g^2}{32\pi^3M}\frac{\del F^2_1}{\del T} \\
      \frac{dg}{dT}&=&3g\frac{\gamma}{T}-\frac{g^3}{32\pi^3}
      \left(-\frac{\del F^0_1}{\del T} +g\frac{d\sigma}{dT}
      (\inv{4M^2}-\frac{\del F^0_1}{\del M^2})\right)
\eea
These equation are  solved by computer. The result for
some different values of zero temperature mass and coupling
are shown in Fig.1 and 2. The effect of solving the
TRGE instead of taking only the naive one-loop result is more
important for large couplings.

It has some times been argued that an asymptotically free
theory (in the sense that the usual $\beta$-function
starts out negative for small couplings) should behave as a
free theory also at high temperature.
This conclusion is correct if the momentum is scaled at the
same rate as $T$. Such a scheme was studied in Ref.\cite{enqvistk92}.
In this paper the momentum is put to zero and we are thus studying
the properties of a low lying excitation embedded in a hot
heat-bath which is rather an IR limit. In this case the coupling
increases with temperature.
%
%
\section{The $(\phi^4)_4$-model}
\label{phifyra}
Spontaneously broken theories are very important in particle
physics and the prototype is the $\phi^4$-model that we shall
study in 4 dimensions. At low temperatures the field gets an
expectation value but above a critical temperature it is
expected that the symmetry is restored. Usual perturbation
theory is bad close to the critical point where the mass
goes to zero and IR divergences occur. A resummation of an infinite
set of diagram to cure the IR problems was performed for the
massless $\phiff$-model in Ref.\cite{altherr90}. In contrast to the
$\phits$-model the $\phiff$-model is stable at all temperatures
since the potential is bounded from below.
\\ \\
The Lagrangian in $D=4-2\epsilon$ is given by
\bea
\lefq{
      L=\inv{2}(\del_\mu\phi_0)^2+\frac{m^2_0}{2}\phi^2_0
      -\frac{\lambda_0}{4!}\phi^4_0} \non\\
      &=&\inv{2}(\del_\mu\phi)^2-\frac{M^2}{2}\phi^2
      -\frac{\lambda\sigma\me}{3!}\phi^3
      -\frac{\lambda\mte}{4!}\phi^4
      +\frac{Z-1}{2}(\del_\mu\phi)^2\non\\
      & &-\sigma\mme\left(\frac{\lambda_0Z^2\sigma^2\mmte}{6}
      -m^2_0Z\right)\phi-\inv{2}\left(\frac{(\lambda_0Z^2-\lambda\mte)
      \sigma^2\mmte}{2}-(m^2_0Z-m^2)\right)\phi^2\non\\
      & &-\frac{(\lambda_0Z^2-\lambda\mte)
      \sigma\mme}{3!}\phi^3-\frac{(\lambda_0Z^2-\lambda\mte)}{4!}
      \phi^4\ ,
\eea
where $M^2=\frac{\lambda\sigma^2}{2}-m^2$. The calculation
of the $N$-point function is similar to the $\phi^3$ case
and we use the same convention regarding the momentum
subtraction point. We then find
\be
   \Gamma^{(1)}=
      -\sigma\mme(\frac{\lambda_0Z^2\sigma^2\mmte}{6}-m^2_0Z)
      +\frac{\lambda\sigma M^2}{32\pi^2}\left(\inv{\epsilon}
      +1-\gamma-\ln (\frac{M^2}{4\pi\mu^2})\right)
      -\frac{\lambda\sigma}{4\pi^2}F^2_1(T,M^2)\ .
\ee
Higher $N$-point functions at zero external momenta can be
computed by taking successive derivatives with respect to
$\sigma$ (remembering the $\sigma$ dependence in $M^2$). Since
we also need the 2-point function at non-zero external momentum
we compute it for $p_\mu=(Q,0,0,0)$ and let $Q$ go to zero at
the end. The $N$-point functions can also be calculated from the
diagrammatic expansion in the shifted theory
with the same result. It may seem
unnecessary to compute the diagrams which are of higher order in
$\lambda$ but they are, on the other hand, more IR divergent.
Using the tree level relation $\lambda\sigma^2=3M^2$ at the local
minimum in the broken phase, we find that
they are all of the same leading order in $\lambda T/M$.  The loop
expansion is an expansion in $\hbar$ and not in $\lambda$ so
we can consistently keep all order terms in $\lambda$. In the unbroken
phase the $\lst$ terms do not contribute.

The other functions needed for renormalization are
\bea
\lefq{
      \Gamma^{(2)}(T,\sigma,M,\lambda)
      =-(\frac{\lambda_0Z^2\sigma^2}{2}-m^2_0Z)
      +\frac{\lambda (M^2+\lambda\sigma^2)}{32\pi^2}
      \left(\inv{\epsilon}-\gamma-\ln (\frac{M^2}{4\pi\mu^2})\right)
      }\non\\
      & &+\frac{\lambda M^2}{32\pi^2}
      -\frac{\lambda}{4\pi^2}F^2_1(T,M^2)
      -\frac{\lambda^2\sigma^2}{4\pi^2}\frac{\del F^2_1}{\del M^2}\ ,\\
\lefq{
      \frac{\del}{\del p^2_0}\Gamma^{(2)}(p=0,T,\sigma,M,\lambda)=
      Z+\frac{\lambda^2\sigma^2}{32\pi^2}
      \left(\inv{6M^2}+F^2_5(T,M^2)\right) \ ,}\\
\lefq{
      \Gamma^{(4)}(p_i=0,T,\sigma,M,\lambda)=
      -\lambda_0Z^2+\frac{3\lambda^2}{32\pi^2}
      \left(\inv{\epsilon}-\gamma-\ln (\frac{M^2}{4\pi\mu^2})\right)
      }\non\\
      & &-\frac{3\lambda^3\sigma^2}{16\pi^2M^2}
      +\frac{\lambda^4\sigma^4}{32\pi^2M^4}
      -\frac{\lambda^2}{4\pi^2}
      \left(3\frac{\del F^2_1}{\del M^2} + 6\lst
      \frac{\del^2F^2_1}{\del M^4} +(\lst)^2
      \frac{\del^3 F^2_1}{\del M^6}\right)\ .\non\\
\eea
\\ \\
A finite expression for the effective potential is obtained from the
tadpole using the RC at $\sr$
\bea
\label{V4}
\lefq{
      V(\sigma)=}\non\\
      & &-\frac{\sigma^2}{2}\left(\mr+\frac{\lr}{4\pi^2}
      (F^2_1+\lstr\frac{\del F^2_1}{\del M^2})-\right.\non\\
      & &\left.\frac{\lambda^2_R\sigma^2_R}{8\pi^2}\left[\frac{3\lstr}{4\mmr}
      -\frac{(\lstr)^2}{8M^4_R}+3\frac{\del F^2_1}{\del M^2}
      +6\lstr\frac{\del^2 F^2_1}{\del M^4}+(\lstr)^2
      \frac{\del^3 F^2_1}{\del M^6}\right]\right)\non\\
      & & +\frac{\sigma^4}{24}\left(\lr-\frac{\lr^2}{4\pi^2}
      \left[\frac{3\lstr}{4\mmr}
      -\frac{(\lstr)^2}{8M^4_R}+3\frac{\del F^2_1}{\del M^2}
      +6\lstr\frac{\del^2 F^2_1}{\del M^4}+(\lstr)^2
      \frac{\del^3 F^2_1}{\del M^6}\right]\right)\non\\
      & &+\frac{M^4}{64\pi^2}(\ln(\frac{M^2}{\mmr})-\inv{2})
      -\inv{6\pi^2}F^4_1(T,M^2)-\frac{\lambda^2_R\sigma^2}{128\pi^2}
      (2\sigma^2-\sigma^2_R)
\eea
where all functions $F^m_n$ without explicit arguments should
be evaluated at $(\tr,\mmr)$.  But $\tr$ can, of course, be chosen
to be equal to $T$ when the TRGE are solved. When deriving the expression for
$V(T,\sigma)$  the RC in \Eqref{RC1} was not used to eliminate
the infinities. Therefore, the minimum of $V(T,\sigma)$ is not
automatically at $\sr$ as it was in the $(\phi^3)_6$-theory. However,
if $\sr(\tr)$ satisfies the TRGE and $\Gamma^{(1)}$ is TRG
invariant (i.e. satisfies \Eqref{TRGE}) we have
\be
      \frac{d}{d\tr}\Gamma^{(1)}(\tr,\sr;\sr,\mmr,\lr,\tr)=0\ ,
\ee
which ensures that the minimum is at $\sr$ for all $T$ if we
choose $T=\tr$.
The one-loop tadpole is not TRG invariant so we determine $\sr$
through the minimization of $V(T,\sigma)$ instead of
solving the TRGE in order to get $\mmr$ as the mass at the
minimum in our approximation.
\\ \\
The TRGE become considerably more complex in the broken phase
of the $(\phi^4)_4$-theory, but it is straightforward to
derive them and we get
\bea
      0&=&\frac{\lambda\sigma}{4\pi^2}\frac{\del F^2_1}{\del T}
      +\frac{d\sigma}{dT}M^2 \ ,\\
      \frac{\gamma}{T}&=&-\frac{\lambda}{32\pi^2}\left(
      \frac{\lst}{2}\frac{\del F^2_5}{\del T} + \lambda\sigma
      \frac{d\sigma}{dT}\left[\inv{6M^2}-\frac{\lst}{12M^4}
      +F^2_5+\frac{\lst}{2}\frac{\del F^2_5}{\del M^2}\right]\right)\ , \\
      \frac{dM}{dT}&=&M\frac{\gamma}{T}+\frac{\lambda\sigma}{2M}
      \frac{d\sigma}{dT}-\frac{\lambda}{8\pi^2 M}\left(
      -\frac{\del F^2_1}{\del T} - \lst\frac{\del^2 F^2_1}
      {\del T\del M^2}\right.\non\\
      &+ & \left.  \lambda\sigma\frac{d\sigma}{dT}
      \left[\frac{5\lst}{8M^2}-\frac{(\lst)^2}{8M^4}+5\lst
      \frac{\del^2 F^2_1}{\del M^4}+(\lst)^2\frac{\del^3 F^2_1}
      {\del M^6}\right]\right)\ ,\\
      \frac{d\lambda}{dT}&=&4\lambda\frac{\gamma}{T}
      -\frac{\lambda^2}{4\pi^2}\left(-\left[3\frac{\del^2 F^2_1}
      {\del T\del M^2}+6\lst\frac{\del^3 F^2_1}{\del T\del M^4}
      +(\lst)^2\frac{\del^4F^2_1}{\del T\del M^6}\right]\right.\non\\
     &+ &\left.   \lambda\sigma\frac{d\sigma}{dT}\left[-\frac{15}{8M^2}
      +\frac{5\lst}{4M^4}-\frac{(\lst)^2}{4M^6}
      -15\frac{\del^2 F^2_1}{\del M^4} -10\lst
      \frac{\del^3 F^2_1}{\del M^6}-(\lst)^2
      \frac{\del^4 F^2_1}{\del M^8}\right]\right)\ .\non\\
\eea
The unbroken phase, $\sigma=0$, has been studied earlier in
Ref.\cite{fujimotoiny86} and I give the result for completeness.
The TRGE simplify to
\bea
\label{ubtrge}
      \frac{dM}{dT}&=&\frac{\lambda}{8\pi^2 M}
      \frac{\del F^2_1}{\del T}\ ,\non\\
      \frac{d\lambda}{dT}&=&\frac{3\lambda^2}{4\pi^2}
      \frac{\del^2 F^2_1}{\del T\del M^2}\ .
\eea
It is the ratio $T/M$ that defines the IR limit and if we expand
the right hand side of \Eqref{ubtrge} in $T/M$ we get
\bea
\label{ubir}
      \frac{dM}{dT}&=&\frac{\lambda T}{24 M}\ ,\non\\
      \frac{d\lambda}{dT}&=&-\frac{3\lambda^2}{16\pi M}\ .
\eea
For a constant coupling we would get back the usual formula in
\Eqref{m2=t2} but we see from \Eqref{ubir} that $\lambda(T)$
decreases and thus changes the large $T$ behaviour of $M(T)$.
The asymptotic solutions are
\be
      M(T)\simeq\frac{4\pi T}{9\ln(T/T_0)}\ ,\quad
      \lambda(T)\simeq\frac{128\pi^2}{27\ln^2(T/T_0)}\ .
\ee
It can be obtained by guessing that $M\ln T/T$ and $\lambda\ln^2 T$
approaches non zero constants as $T\rightarrow\infty$ and proving
it using \Eqref{ubir}. The constant $T_0$ is introduced for
dimensional reasons and is determined by the initial conditions.
It must, therefore, by of the order of $M(0)$. This analysis
gives another asymptotic $T$ dependence of the effective
4-point coupling $\lambda(T)$ than the one given in Ref.\cite{fendley87}
where the ordinary renormalization group (in $\mu$) was used.
Like the case of the $\phits$-model a different behaviour is expected if
$p\simeq T$ instead of $p=0$ as in this paper.

A two-loop calculation for the $O(N)$-symmetric model in the
limit $N\rightarrow\infty$ was carried out in Ref.\cite{funakubos87}
where it was found that the two-loop effects are important.
%
\section{Conclusions}
\label{concl}
Self-consistent renormalization conditions at zero momentum
together with the
temperature renormalization group equations (TRGE) of
Ref.\cite{matsumotonu84} has been used to improve the IR
behaviour of the scalar $(\phi^3)_6$- and $(\phi^4)_4$-models.
The renormalization condition replaces the gap equation which
resums an infinite set of diagrams and the TRGE  determine
the running of the mass and coupling with temperature.
\\ \\
The $\tr$-renormalization scheme in \Eqref{RC1}-\Eqref{RC4} is not the
only possible scheme. One can use the gap equation (\Eqref{gap}) as
a renormalization prescription (called the $T$-scheme
in Ref.\cite{matsumotonu84}). For an unbroken theory it takes the form
\be
\label{Tsch}
      {\rm Re}\Gamma^{(2)}(p,T,M(T),g(T))|_{p(Q_R)}=
      p^2(Q_R)-M^2(T)\ ,\quad {\rm etc.}
\ee
Since the exact vertex functions should be independent of the
renormalization prescription (up to a finite wavefunction renormalization
factor) the $M(T)$ obtained from \Eqref{Tsch} must be the same as the
$M(T)$ computed from the TRGE. It is, however, not true to finite
order in perturbation theory.
A careless use of the gap equation can lead to temperature
dependent infinities (as pointed out in Ref.\cite{fendley87})
but with the analysis in this paper no such divergences occur.
\\ \\
{}From the numerical solutions of the TRGE we see in the figures
that for small
zero temperature coupling constants the corrections to the
unrenormalized calculations of the critical temperature $T_{cr}$
and the mass $M(T)$ are small. The fact that the corrections in the
$\phiff$-model are larger than in the $\phits$-model for
$g=\lambda$ is due to a difference in the numerical factors
in the TRGE. For zero temperature values of
$g$ and $\lambda$ equal to 1 and 10, respectively,
$T_{cr}$ and the behaviour of $M(T)$ are altered,
but it is $g(T)$ and $\lambda(T)$
that get the most drastic corrections.
The increase in $T_{cr}$ can be explained by the decrease in $g$
for the  $\phits$-model,
and vice versa for the $\phiff$-model.

Since it is the coupling constants that get the largest corrections
it would be interesting to use the TRGE to study a model with a
first order phase transition (e.g. the electroweak theory).
The height of the barrier between the coexisting phases depends
strongly on the coupling constant.
\\ \\
When we use the renormalization conditions to express the vertex
functions in terms of finite quantities all the dependence of
the arbitrary scale $\mu$ disappears. Instead we get a logarithmic
dependence of the $\sigma$ through $M^2(\sigma)$ (see \Eqsref{V3}{V4}).
The effective potential is not to be trusted when these logarithms are
large. A renormalization group in $\sigma$, similar to the one in
 $T$, should be used to improve
the potential for $\sigma$ far away from the minimum. The value
of $\sr$, at which we renormalize, is arbitrary, but by choosing the
minimum we get $M(\tr)$ to be the mass of the lowest physical excitation.
At the minimum the logarithmic terms are zero and the form of the
potential reliable. We can, therefore, follow the temperature dependence
of $\sr$. In the $(\phi^4)_4$-model
we find that it goes to zero at the same temperature as the
mass. Thus, the phase transition is of second order.
\\ \\
In the high temperature limit of the $(\phi^4)_4$-model we find that the
ratio $T/M$, which enters as a possible IR divergence in the
perturbation expansion, goes like $9\ln T/4\pi$ for large $T$
because also the mass grows with $T$. But, for a given $N$-point
function the IR factor is actually $(\lambda T/M)^V$
($V$ is the number of vertices in a diagram), apart from an overall
factor that depends only on $N$ and not on $V$. In UV divergent
diagrams there is an extra factor $T/M$.
If one also takes into account the $T$ dependence
of $\lambda$ we actually find that
\be
      \frac{\lambda T}{M}\simeq\frac{32\pi}{3\ln(T/T_0)}\ ,
\ee
and it thus {\it decreases} at high temperature. This indicates that
a perturbation expansion could be possible in the large $T$ limit
if only the renormalized parameters are used, but the extra factors
of $T/M$ from UV divergent diagrams may destroy the convergence.
See Ref.\cite{funakubos87} for the effects of higher loops.
Note that the coupling is evaluated at zero external momenta. If
we put $p/T=const.$, as suggested in Ref.\cite{enqvistk92},
we expect to get back the result from the usual renormalization
group analysis. A renormalization group in $p_0$ and $\Abs{\vec{p}}$
should be used to study the momentum dependence in more detail.
\\ \\
I wish to thank K. Farakos for discussions and M. Sakamoto
for drawing my attention to Ref.\cite{funakubos87}.
%

%
%
\
\newpage
\section*{\centering Figure Captions}
{\bf Figures}
Numerical calculation of the temperature dependence of $M(T)$,
$g(T)$ and $\lambda(T)$ in the $\phits$- and $\phiff$-models.
The temperature, mass and couplings are rescaled to make it
easier to compare with the unrenormalized one-loop result
in which cases the curves are straight lines.
\begin{enumerate}
  \item The mass in the $\phits$-model.
  \item The coupling in the $\phits$-model.
  \item The mass in the $\phiff$-model.
  \item The coupling in the $\phiff$-model.
\end{enumerate}

\begin{thebibliography}{99}
%
\bibitem{dinelhll92}
      M. Dine, R. Leigh, P. Huet, A. Linde and D. Linde,
      ``{\sl Towards the theory of the electroweak
      phase transition }", SLAC-PUB-5741, SCIPP-92-07, SU-ITP-92-7
%
\bibitem{dolanj74}
      L. Dolan and R. Jackiw,
      ``{\sl Symmetry behavior at finite temperature }",
      \prd{9}{1974}{3320}
%
\bibitem{carrington91}
      M. E. Carrington,
      ``{\sl The effective potential at finite temperature in
      the standard model }", TPI-MINN-91/48-T
%
\bibitem{matsumotonu84}
      H. Matsumoto, Y. Nakano and H. Umezawa,
      ``{\sl Renormalization group at finite temperature }",
      \prd{29}{1984}{1116}
%
\bibitem{fujimotoy88b}
      Y. Fujimoto and H. Yamada, ``{\sl Finite-temperature
      renormalization group equation in QCD. II }",
      \pl{B200}{1988}{167}
%
\bibitem{nakkagawany88}
      H. Nakkagawa, A. Ni\'{e}gawa and H. Yokota,
      ``{\sl Non-Abelian gauge couplings at finite temperature
      in the general covariant gauge }", \prd{38}{1988}{2566}
%
\bibitem{freires91}
      F. Freire and C. R. Stephens,
      ``{\sl What can we learn from the finite temperature
      renormalization group? }", Imperial/TP/91/92/15
%
\bibitem{altherrgp91}
      T. Altherr, T. Grandou and R. D. Pisarski,
      ``{\sl Thermal instability in $(\phi^3)_6$}",
      \pl{B271}{1991}{183}
%
\bibitem{fujimotoy88}
      Y. Fujimoto and H. Yamada, ``{\sl A supplementary
      remark of finite temperature perturbation }",
      \zp{C37}{1988}{265}
%
\bibitem{altherr90}
      T. Altherr, ``{\sl Infrared problem in $g\phi^4$ theory
      at finite temperature }", \pl{B238}{1990}{360}
%
\bibitem{fujimotoiny86}
      Y. Fujimoto, K. Ideura, Y. Nakano and H. Yoneyama,
      ``{\sl The finite temperature renormalization group
      equation in $\lambda\phi^4$ theory }",
      \pl{B167}{1986}{406}
%
\bibitem{funakubos87}
      K. Funakubo and M. Sakamoto,
      ``{\sl Higher order contribution in finite temperature
      renormalization group }",
      \pl{B186}{1987}{205}
%
\bibitem{fendley87}
      P. Fendley, ``{\sl The effective potential
      and the coupling constant at high temperature }",
      \pl{B196}{1987}{175}
%
\bibitem{enqvistk92}
      K. Enqvist and K. Kainulainen,
      ``{\sl The running coupling constant at finite
      temperature }'', \zp{C53}{1992}{87}
%
\end{thebibliography}
\end{document}